\documentclass[12pt,a4]{article}
\begin{document}
\baselineskip 11pt
\pagestyle{empty}
\begin{titlepage}

\rightline{IC/97/208}
\vspace{2.0 truecm}
\begin{center}
\begin{Large}
{\bf $Z^\prime$ and anomalous gauge coupling effects at LEP2 \\
and their separation}

\end{Large}

\vspace{2.0cm}
{\large A. A. Babich\hskip 2pt\footnote{Permanent address: Department of
Mathematics, Polytechnical Institute, Gomel, 246746 Belarus. E-mail:
BABICH@GPI.GOMEL.BY}}
\\[0.3cm]
International Centre for Theoretical Physics, Trieste, Italy

\vspace{5mm}

{\large  A. A. Pankov\hskip 2pt\footnote{Permanent address: 
Department of Physics, Polytechnical Institute, Gomel, 
246746 Belarus. E-mail: PANKOV@GPI.GOMEL.BY}
}\\[0.3cm]
International Centre for Theoretical Physics, Trieste, Italy\\
Istituto Nazionale di Fisica Nucleare, Sezione di Trieste, Trieste,
Italy

\vspace{5mm}

{\large  N. Paver\hskip 2pt\footnote{Also supported by the Italian
Ministry of University, Scientific Research and Technology (MURST).}
}\\[0.3cm]
Dipartimento di Fisica Teorica, Universit\`{a} di Trieste,
Trieste, Italy\\
Istituto Nazionale di Fisica Nucleare, Sezione di Trieste, Trieste,
Italy
\end{center}
\vspace{2.0cm}

\begin{abstract}
\noindent
In the case of the process $e^+e^-\to\mu^+\mu^-$, we study the sensitivity    
to either $Z^\prime$ or anomalous gauge coupling effects  
of two new observables, $\sigma_+$ and $\sigma_-$, conveniently defined for 
such kind of analysis. We discuss general properties of the deviations of these 
observables from the Standard Model predictions, and derive the 
model-independent bounds on the 
relevant parameters allowed by present and foreseeable LEP2 data.  
We also discuss the possibility of separating the two kinds of 
effect. 

 \vspace*{3.0mm}

\noindent
\end{abstract}
\end{titlepage}
\pagestyle{plain}
\setlength{\baselineskip}{1.3\baselineskip}

\section{Introduction}
The existence of a heavy neutral gauge boson $Z^\prime$ is theoretically
motivated by most extended electroweak scenarios attempting to overcome the 
conceptual difficulties of the Standard Model \cite{rizzo}. The knowledge of 
the $Z^\prime$ parameters, such as the mass and couplings to ordinary fermions, 
would be essential in order to test these extended theories. Numerous 
strategies to evidence manifestations of the $Z^\prime$ in experiments at high 
energy $e^+e^-$ and hadronic colliders have been developed.    
Due to the present  
`direct' lower mass limit from the Tevatron, $M_{Z^\prime}>{\cal O}(500)$ GeV 
\cite{Tevatron}, only `indirect' manifestations of the $Z^\prime$ can be 
searched for at the energy of LEP2. The preferred channel in this case 
should be the electron-positron annihilation into fermion pairs, where 
the $Z^\prime$ contributes already at the Born level \cite{lep2}. In this 
regard, particularly convenient should be the annihilation into lepton pairs 
\begin{equation} e^++e^-\to {\bar l}+l \qquad \mbox{($l=\mu$, $\tau$)}
\label{proc}\end{equation}
where, assuming lepton universality, the number of $Z^\prime$ parameters is 
significantly reduced, leading to characteristic, model-independent potential
signals. These effects manifest themselves as deviations of observables from the 
Standard Model (SM) predictions, to be compared to the experimental data, not 
only as regards the size of the signals, but also their characteristic signs 
and energy dependences. In the case of no observed signal within the
experimental accuracy, the result of the analysis can be expressed as bounds on the 
$Z^\prime$ parameters to a conventionally specified confidence level.\par 
Similar effects in process (\ref{proc}) can arise from the independent physical 
mechanism represented by anomalous gauge boson couplings, that can contribute 
even at the tree level \cite{Hagiwara}. Clearly, in the case of no observed 
effect, upper limits on the anomalous gauge couplings can be obtained from the 
data. On the other hand, it would be desirable to have some criterion to 
distinguish the two sources of such nonstandard effects, in the case these were 
observed.   
\par 
The aim of this paper is to present a discussion of the general properties of 
the deviations from the SM predictions due to the $Z^\prime$ and to 
the anomalous gauge couplings, at the energies of LEP2. In particular, 
some criterion to distinguish the two kinds of signal is worked out. Moreover, 
the constraints on the relevant parameters are derived either from the present 
LEP2 data and in the case of the foreseeable ones.

\section{Manifestations of $Z^\prime$}
We start from the general neutral-current interaction including the $Z^\prime$: 
\begin{equation} -L_{NC}=eJ_{\gamma}^{\mu} A_{\mu}+g_ZJ_{Z}^{\mu} Z_{\mu}
+g_{Z^\prime}^{\mu}J_{Z^\prime}^{\mu} Z_{\mu}^{\prime}, \end{equation}
where $e=\sqrt{4\pi\alpha_{e.m.}}$; $g_Z=e/s_W c_W$ (with $s_W^2=1-c_W^2\equiv 
\sin^2\theta_W$) and $g_{Z^\prime}$ are the $Z$ and $Z^\prime$ gauge couplings, 
respectively. The currents can be written as
\begin{equation} 
J_i^{\mu}=\sum_f{\bar\psi}_f\gamma^{\mu}(L_i^f P_L+R_i^f P_R)\psi_f
=\sum_f{\bar\psi}_f\gamma^{\mu}(V_i^f-A_i^f\gamma_5)\psi_f.\label{currents}
\end{equation}
Here, $i=\gamma, Z, Z^\prime$ and $P_{L,R}=(1\mp\gamma_5)/2$ are projectors 
onto the left- and right-handed electron helicity states, so that
$R_i^f=V_i^f-A_i^f$ and $L_i^f=V_i^f+A_i^f$.
The Standard Model (SM) couplings are 
\begin{equation}
V_{\gamma}^{f}=Q_f;\qquad A_{\gamma}^f=0;\qquad V_Z^{f}=\frac{I_{3L}^f}{2}-
Q_f s_W^2;\qquad A_Z^f=\frac{I_{3L}^{f}}{2}.\label{smcoupl}\end{equation}
\par 
In general, the unpolarized differential cross section for electron-positron 
annihilation into a fermion-antifermion pair, $e^+e^-\to \bar{f} f$, can 
be written in Born approximation including $\gamma$, $Z$ and $Z^\prime$ 
exchanges, as:
\begin{equation}
\frac{d\sigma_{ff}}{d\cos\theta}
=\frac{\pi\alpha_{e.m.}^2}{2s}
\left[(1+\cos^2\theta)\ F_1 +2\cos\theta\ F_2\right], \label{cross} 
\end{equation}
where $\theta$ is the angle between the initial electron and the outgoing
fermion. 
\par
For our purposes, it is convenient to split $F_1$ and $F_2$ as:
\begin{equation}
F_1=F_1^{SM}+\Delta F_1, \qquad F_2=F_2^{SM}+\Delta F_2.
\label{f}\end{equation}
In Eq.~(\ref{f}):
\begin{eqnarray}
F_1^{SM}&=&Q^2_eQ^2_f+2\ Q_ev_eQ_fv_f {\rm Re}\chi_Z
+(v^2_e+a^2_e)(v^2_f+a^2_f)|\chi_Z|^2,\nonumber \\
F_2^{SM}&=&2\ Q_ea_eQ_fa_f{\rm Re}\chi_Z
+4\ v_ea_ev_fa_f|\chi_Z|^2, \label{ff}\end{eqnarray} 
and the deviation due to the $Z^\prime$ is
\begin{eqnarray}
\Delta F_1&=&2\ Q_ev'_eQ_fv'_f{\rm Re}\chi_{Z'}
+(v_e'{}^2+a_e'{}^2)(v_f'{}^2+a_f'{}^2)|\chi_{Z'}|^2 \nonumber \\
& &
+2\ (v_ev'_e+a_ea'_e)(v_fv'_f+a_fa'_f){\rm Re}(\chi_{Z}\chi^*_{Z'}),
\nonumber \\
\Delta F_2&=&2\ Q_ea'_eQ_fa'_f{\rm Re}\chi_{Z'}
+4\ v'_ea'_ev'_fa'_f|\chi_{Z'}|^2\nonumber  \\
& &
+2(v_ea'_e+v'_ea_e)(v_fa'_f+v'_fa_f){\rm Re}(\chi_{Z}\chi^*_{Z'}).
\nonumber
\label{df}\end{eqnarray}
Here, the gauge boson propagators are $\chi_V=s/(s-M^2_V+iM_V\Gamma_V)$, 
$V=Z$, $Z'$, and the fermionic coupling constants are normalized as:
\begin{equation}
v_f=\frac{g_Z}{e}V_Z^f,\qquad a_f=\frac{g_Z}{e}A_Z^f,\qquad
v_f^{\prime}=\frac{g_{Z^\prime}}{e}V_{Z^\prime}^f \qquad 
a_f^{\prime}=\frac{g_{Z^\prime}}{e}A_{Z^\prime}^f.\label{coupli}
\end{equation}
The total width $\Gamma_{Z'}$ can be assessed from a sum over
the partial widths:
\begin{equation}
\Gamma_{Z'}^{ff}=\frac{\alpha_{e.m.}M_{Z'}}{3}
\sqrt{1-4m^2_f/M_{Z'}^2}\left[
{v_f^\prime}^2+{a_f^\prime}^2+2m^2_f/M_{Z'}^2\hskip 2pt
({v_f^\prime}^2-2{a_f^\prime}^2)\right].\label{width}
\end{equation}

The leptonic channel (\ref{proc}) has the nice phenomenological feature that,  
assuming leptonic $e-l$ universality, only three independent nonstandard 
parameters are needed: the couplings $v_{l}^{\prime}$, $a_{l}^{\prime}$
and the mass $M_{Z^\prime}$. More parameters would be needed in the description 
of $e^+e^-\to{\bar q}q$.\par  
Concentrating on process (\ref{proc}) at LEP2 energies with $l=\mu$, the two 
conventionally used observables are the total cross section
\begin{equation}
\label{crosstot}
\sigma_{\mu\mu}=\int\limits_{-1}^{1}
\frac{d\sigma_{\mu\mu}}{d\cos\theta} d\cos\theta
=\sigma_{\rm pt}F_1 ,
\end{equation}
with $\sigma_{\rm pt}\equiv\sigma(e^+e^-\to\gamma^\ast\to\mu^+\mu^-)=
(4\pi\alpha_{e.m.}^2)/(3s)$, and the forward-backward asymmetry 
\begin{equation}
\label{AFB}
A_{\rm FB}=\frac{\sigma_{\rm FB}}{\sigma_{\mu\mu}}
\equiv\frac{\sigma_{\mu\mu}^{\rm F}-\sigma_{\mu\mu}^{\rm B}}{\sigma_{\mu\mu}} 
=3F_2/4F_1,
\end{equation}
where the forward and backward cross sections are defined by 
$\sigma_{\mu\mu}^{\rm F}
=\int_{0}^{1}(d\sigma_{\mu\mu}/d\cos\theta)d\cos\theta$ and  
$\sigma_{\mu\mu}^{\rm B}
=\int_{-1}^{0}(d\sigma_{\mu\mu}/d\cos\theta)d\cos\theta$, respectively. 
\par 
As it will appear in the sequel, in order to identify $Z^\prime$ 
effects in process (\ref{proc}) it may be convenient to consider,   
alternatively to $A_{\rm FB}$, just the difference $\sigma_{\rm FB}$ between 
the forward and backward cross sections. 
\par 
To discuss the distinctive features of nonstandard physical  
effects induced either by the exchange of a $Z'$ or by anomalous gauge 
couplings, we introduce also the new observables $\sigma_+$ and $\sigma_-$ 
defined as the differences of cross sections integrated in suitable ranges of 
$\cos\theta$:
\begin{eqnarray}
\label{sigma+}
\sigma_+&\equiv&\left(\int_{-z^*}^1-\int_{-1}^{-z^*}\right)
\frac{d\sigma_{\mu\mu}}{d\cos\theta}\ d\cos\theta, \\
\label{sigma-}
\sigma_-&\equiv&\left(\int_{-1}^{z^*}-\int_{z^*}^1\right)
\frac{d\sigma_{\mu\mu}}{d\cos\theta}\ d\cos\theta,
\end{eqnarray}
where $z^*>0$ is determined from the condition that the kinematical
coefficients multiplying $F_1$ and $F_2$ in Eq.~(\ref{cross}) be the same 
after integration over $\cos\theta$. The condition is 
\begin{equation}
\int_{-z^*}^{z^*}(1+\cos^2\theta)\ d\cos\theta
=\left(\int_{z^*}^1-\int_{-1}^{-z^*}\right)\ 2\cos\theta\ d\cos\theta,
\end{equation}
and gives the equation for $z^*$ 
\begin{equation}
\frac{1}3 z^*{}^3+z^*{}^2+z^*-1=0,
\end{equation}
whose solution is $z^*=2^{2/3}-1=0.5874$, corresponding to
$\theta^*=54^\circ$. In the case of a reduced kinematical range of the 
scattering angle such as, e.g., $\vert\cos\theta\vert<c$, one derives 
${z^*=(1+3c)^{1/3}-1}$.  
One can write $\sigma_+$ and $\sigma_-$ in terms of $F_1$ and $F_2$ as
\begin{equation}
\label{sigmapm}
\sigma_{\pm}=\sigma^*_{\rm pt}\ (F_1\pm F_2),
\end{equation}
where, to a very good precision:
\begin{equation}
\sigma^*_{\rm pt}
\equiv\frac{3}{4}\left(1-z^*{}^2\right)\sigma_{\rm pt}
=0.5\,\sigma_{\rm pt}.
\label{pt*}
\end{equation}
The new, independent, observables $\sigma_{\pm}$ 
are simple combinations of the conventional ones $\sigma_{\mu\mu}$ and 
$A_{\rm FB}$. Indeed, from Eqs.~(\ref{crosstot}), (\ref{AFB}), (\ref{sigmapm}) 
and (\ref{pt*}):
\begin{equation}
\label{relation}
\sigma_{\pm}=0.5\,\sigma_{\mu\mu}
\left(1\pm\frac{4}{3}A_{\rm FB}\right)=
0.5\,\left(\sigma_{\mu\mu}\pm\frac{4}{3}\sigma_{\rm FB}\right),
\end{equation}
so that $\sigma_+$ and $\sigma_-$ can be measured either directly according to 
Eqs.~(\ref{sigma+}) and (\ref{sigma-}) or indirectly by means of
$\sigma_{\mu\mu}$ and $A_{\rm FB}$.

The previous formulae for the differential cross section (\ref{cross}), 
as well as those for all of the other observables, 
are still valid to a very good (improved Born) approximation which includes 
one-loop electro-weak radiative corrections, with the following
replacements \cite{Hollik}, \cite{Altarelli2}:
\begin{equation}\begin{array}{lll}
{\displaystyle{\alpha_{e.m.}\Rightarrow\alpha(M_Z^2)}};
&{\displaystyle{ v_l\Rightarrow
\frac{1}{\sqrt{\kappa}}\left(I^l_{3L}-2Q_e\sin^2\theta^{\rm eff}_W\right)}}; 
&
{\displaystyle{a_l\Rightarrow\frac{I^l_{3L}}{\sqrt{\kappa}}}}\\
{\displaystyle{\sin^2\theta_W\Rightarrow
\sin^2\theta^{\rm eff}_W}};&
{\displaystyle{\sin^2(2\theta^{\rm eff}_W)
\equiv\kappa=\frac{4\pi\alpha(M_Z^2)}{\sqrt{2}G_{\rm F}\, M_Z^2\rho}}}; 
&{\displaystyle{ 
\rho\approx 1+\frac{3 G_{\rm F} m^2_t}{8\pi^2\sqrt{2}}}}, 
\end{array}
\label{impborn} \end{equation}
where only the main contribution to $\rho$, coming from the {\it top} mass,
has been taken into account. This parameterization uses the best known SM 
parameters $G_{\rm F}$, $M_Z$, and $\alpha(M^2_Z)$. Moreover, the energy
dependence of the $Z$-width should be accounted for by the replacement of the 
$Z$ propagator:
\begin{equation}
\label{prop}
\chi_Z(s)\Rightarrow\frac{s}{s-M_Z^2+i(s/M_Z^2)M_Z\Gamma_Z}.
\end{equation}
In our analysis, the improved Born approximation with $m_t=175$~GeV and
$m_H=300$~GeV is used. 
\par 
While the numerical results are obtained from the exact formulae, 
for the sake of a simplified presentation of the analysis it is convenient to 
make a few, harmless, assumptions: 
\par\noindent
{\it i}) since the typical lower bound for the $Z^\prime$ boson mass,
is much larger than the 
LEP2 energy, it suffices to take into account $Z^\prime$ interference effects
only, the pure $Z^\prime$ exchange contributions being negligible;
\par\noindent
{\it ii}) since in the SM
$\vert v_l\vert\ll\vert a_l\vert<1$, in the expressions of the 
deviations from the SM due to the $Z^\prime$ we neglect $v_l$ with respect to 
$a_l$. In addition, one can neglect the imaginary part of the $Z^\prime$ boson 
propagator.
\par 
In the approximations {\it i}) and {\it ii}), the deviation of the cross 
section from the SM prediction,  
${\Delta\sigma_{\mu\mu}=\sigma_{\mu\mu}-\sigma_{\mu\mu}^{SM}}$, reads 
\begin{equation}
\label{Delsigma}
\Delta\sigma_{\mu\mu}
\approx 2\sigma_{\rm pt}\left(v_l'{}^2+a_l'{}^2\,a_l^2\ 
{\rm Re}\ \chi_Z\right)\chi_{Z'},
\end{equation}
where the two terms represent the $\gamma-Z^\prime$ and $Z-Z^\prime$ 
interference, respectively, and $v_l'$, $a_l'$ are the leptonic vector and 
axial-vector couplings to the $Z^\prime$ defined in Eq.~(\ref{coupli}). 
\par 
In Eq.~(\ref{Delsigma}), only squares of leptonic couplings appear, so that     
one has the general property that the sign of the deviations of the cross 
section from the SM  prediction due to the $Z^\prime$ is definitely negative 
at the LEP2 energy, because both terms  within the round brackets are 
positive. From Eq.~(\ref{df}), it is clear that this is independent of the 
simplifying assumption {\it ii}) above. Also, such sign definiteness holds 
only for process (\ref{proc}), assuming universality between initial and final 
leptons $\mu$ or $\tau$, and not for the the process $e^+e^-\to q{\bar q}$, 
where additional assumptions on the fermion coupling constants would be needed. 
\par 
The deviation of $\sigma_{\rm FB}$ in (\ref{AFB}) from the SM prediction, 
${\Delta\sigma_{\rm FB}= 
 \sigma_{\rm FB}-\sigma_{\rm FB}^{\rm SM}}$, has a similar expression:
\begin{equation}
\Delta\sigma_{\rm FB}\approx
\frac{3}{2}\,\sigma_{\rm pt}\left(a_l'{}^2+v_l'{}^2\,a_l^2\,
{\rm Re}\,\chi_Z\right)\chi_{Z'} , 
\label{delsigmafb1}
\end{equation}
and at LEP2 it generally has the same negative sign property 
for arbitrary leptonic $Z^\prime$ couplings as the cross 
section $\sigma_{\mu\mu}$.
Such sign property of $Z^\prime$ effects for these observables 
can be used to attempt a model-independent distinction of the $Z^\prime$ from 
analogous anomalous gauge coupling effects, if some deviations were observed.
\par 
Concerning the forward-backward asymmetry (\ref{AFB}), the deviation 
from the SM prediction is expressed as
\begin{equation}
\label{delafb}
\Delta A_{\rm FB}=A_{\rm FB}-A_{\rm FB}^{\rm SM}
\propto \left[
\left(a_l'{}^2-\frac{4}{3}\, A_{\rm FB}^{\rm SM}\, v_l'{}^2\right)
+\left(v_l'{}^2-\frac{4}{3}\, A_{\rm FB}^{\rm SM}\, a_l'{}^2\right)
a_l^2\, {\rm Re}\chi_Z\right]\chi_{Z'} 
\end{equation}
and, in contrast to the case of $\sigma_{\mu\mu}$ and $\sigma_{\rm FB}$, at  
LEP2 energy it can be either positive or negative, depending on the actual 
values of the $Z^\prime$ couplings.
\par 
The deviation of $\sigma_+$ from the SM prediction has the expression 
\begin{equation}
\label{deltasig+}
\Delta\sigma_{+}=\sigma_{+}-\sigma_{+}^{\rm SM} 
\approx
\sigma_{\rm pt}\,
(1+ a_l^2\,{\rm Re}\ \chi_{Z})\cdot (v'_l{}^2+ a'_l{}^2)\chi_{Z'}.
\end{equation}
Eq.~(\ref{deltasig+}) shows that
the dependence of $\Delta\sigma_{+}$ on the $Z^\prime$ parameters
is characterized by the expression $(v_l^\prime{}^2+ a^\prime_l{}^2)\chi_{Z'}$,
which is a negative quantity at $\sqrt{s}<M_{Z^\prime}$ for arbitrary $Z^\prime$ 
couplings. In addition to the definite negative sign, $\Delta\sigma_{+}$ has 
the property of depending on the total center of mass energy $\sqrt s$ 
{\it via} the factor $(1+ a_l^2\, {\rm Re}\,\chi_{Z})$, which is completely 
determined by well-known SM parameters. Fig.~1a shows the energy dependence of 
the relative deviation $\Delta\sigma_+/\sigma_+^{\rm SM}$. In this numerical 
example, we have considered a `sequential' $Z^\prime$ \cite{Mele} with 
couplings ${v_l^\prime=v_l}$, ${a_l^\prime=a_l}$ and with masses 
$M_{Z^\prime}=500$ GeV 
and $M_{Z^\prime}=700$ GeV. In this figure, one should notice the  
characteristic sign correlations of $\Delta\sigma_+$ at the different energies 
expected from a $Z^\prime$ ($\Delta\sigma_+/\sigma_+^{SM}$ is sequentially 
negative, positive and then negative again, with increasing $\sqrt s$). 
\par    
As regards $\sigma_{-}$, it has in common with $\sigma_+$ the property that 
the deviation from the SM prediction
\begin{eqnarray}
\label{Dsigm}
\Delta\sigma_{-}=\sigma_{-}-\sigma_{-}^{\rm SM} 
\approx
\sigma_{\rm pt}\,
(1- a_l^2\, {\rm Re}\ \chi_{Z})\cdot (v^\prime_l{}^2- a^\prime_l{}^2)
\chi_{Z^\prime}
\end{eqnarray}
has an energy dependence completely determined in terms of SM parameters. 
However, in contrast to $\Delta\sigma_+$, the sign and magnitude of 
$\Delta\sigma_{-}$ is determined by
$(v_l^\prime{}^2-a_l^\prime{}^2)\chi_{Z^\prime}$, for which both the positive
and the negative signs are generally possible. Fig.~1b shows the energy 
dependence of the relative deviation $\Delta\sigma_-/\sigma_-^{\rm SM}$, 
which, similar to $\Delta\sigma_{+}$, has the characteristic sign 
correlations at the different energies. In the case of Fig.~1b, we take a 
$Z^\prime$ with same mass values as in Fig.~1a, but with couplings such that 
${v^\prime_l}^2-{a^\prime_l}^2=\pm({v_l}^2-{a_l}^2)$ (to give an 
example). 
\begin{table}
\begin{center}
\begin{tabular}{|c|c|c|}
\hline
Observables &Sign defined & Energy dependence defined\\ \hline
$\Delta\sigma_{\mu\mu}$   & ${\rm yes}$ & ${\rm no}$ \\
$\Delta\sigma_{{\rm FB}}$ & ${\rm yes}$ & ${\rm no}$ \\
$\Delta A_{{\rm FB}}$     & ${\rm no}$ & ${\rm no}$ \\
$\Delta\sigma_+$          & ${\rm yes}$ & ${\rm yes}$ \\
$\Delta\sigma_-$          & ${\rm no}$ & ${\rm yes}$ \\
\hline
\end{tabular}
\end{center}
\caption{Model-independent properties of deviations induced by a $Z^\prime$ 
at LEP2 energy}
\end{table}
\par
The properties of the deviations $\Delta{\cal O}$ of the observables considered 
above (${\cal O}=\sigma_{\mu\mu}$, 
$\sigma_{{\rm FB}}$, $A_{{\rm FB}}$, $\sigma_\pm$), are summarized in Table~1.
As previously emphasized, these properties of the observables
may be useful to identify (or to discard) the 
$Z^\prime$ as the (indirect) source of some deviations from the SM predictions, 
if experimentally observed. 
\par
Alternatively, it should be interesting to assess 
the sensitivity to the $Z^\prime$ of the observables ${\cal O}$, 
and the corresponding 
potential to constrain $Z^\prime$ parameters within an allowed region, in the 
case no deviations were observed within the expected experimental accuracy. 
We may attempt a model-independent analysis, which accounts either 
for the uncertainty of experimental data currently available from LEP2, 
or for the accuracy foreseeable for the future data with the planned integrated 
luminosity of 500 $pb^{-1}$. To this purpose, it is convenient to adopt the 
general expression of the neutral current interaction of the $Z'$ given, 
e.g., in Refs.~\cite{Leike,p1}, and rewrite (\ref{df}) in terms of the 
`effective' vector and axial-vector couplings
\begin{equation}
\label{coupl}
V_f=V^f_{Z'}\, \sqrt{\frac{g^2_{Z'}}{4\pi}\,
\frac{M^2_Z}{M^2_{Z'}-s}},
\qquad
A_f=A^f_{Z'}\, \sqrt{\frac{g^2_{Z'}}{4\pi}\,
\frac{M^2_Z}{M^2_{Z'}-s}}.
\end{equation}
One of the advantages is that (\ref{coupl}) allows to represent bounds on a 
two-dimensional scatter plot, without making reference to specific values 
of $M_{Z^\prime}$ or $s$.\par        
The sensitivity of the observables $\sigma_\pm$ has been assessed 
numerically by defining a $\chi^2$ function as follows:
\begin{equation}
\label{Eq:chisq}
\chi^2
=\left(\frac{\Delta{\cal O}}{\delta{\cal O}}\right)^2,
\end{equation}
where the uncertainty $\delta{\cal O}$ combines both statistical
and systematic errors. As a criterion to derive allowed regions for
the coupling constants in the case where no deviations
from the SM were observed, and in this way to assess the sensitivity
of process (\ref{proc}) to $V_l$ and $A_l$ through the observables
${\cal O}$, we impose that $\chi^2<\chi^2_{\rm crit}$, where 
$\chi^2_{\rm crit}$ is a number that specifies the desired `confidence' level. 
We take the current integrated luminosities and systematic uncertainty
relevant to the cross section of 
the leptonic process $e^+e^-\to\mu^+\mu^-$ at LEP2, as summarized in Table~2.
\begin{table}
\begin{center}
\begin{tabular}{||l|r|r|r|r||} \hline
$\sqrt{s}$ &161 [GeV] & $\delta_{\rm syst}$ & 172 [GeV]
&$\delta_{\rm syst}$ 
\\ \hline

L3 \cite{l3}         & 10.9 $pb^{-1}$ & 4.0\% & 10.2 $pb^{-1}$   &  4.0\%  \\
OPAL \cite{opal}     & 10.1 $pb^{-1}$ & 2.2\% & 10.3 $pb^{-1}$   &  2.8\%  \\
ALEPH \cite{aleph}   &11.08 $pb^{-1}$ & 2.3\% &10.65 $pb^{-1}$   &  2.3\% \\
DELPHI \cite{delphi} & 9.95 $pb^{-1}$ & 3.5\% & 9.98$ pb^{-1}$   &  3.4\%  \\ \hline
\end{tabular}
\end{center}
\caption{Systematic errors and integrated luminosities  
collected at current experiments at LEP2}
\end{table}
In addition, to assess future possibilities of improving the sensitivity, we 
work out the same example by using the expected final luminosity at LEP2,  
assuming for the systematic uncertainty, in this case, 
$\delta_{\rm syst}=0.5\%$ \cite{lep2}). 
The numerical analysis has been performed by means of the 
program ZEFIT, which has to be used along with ZFITTER \cite{zfitter}.
\par
In Fig.~2 we show the results from $\sigma_\pm$, and compare the bounds 
on the couplings (\ref{coupl}) obtainable 
from current and future experiments at LEP2,
with inputs from Table~2. The contours are derived from the combination
of both observables, ${\sigma_+}$ and $\sigma_-$, and correspond to two 
standard deviations ($\chi^2_{\rm crit}=4$). From the numerical point of 
view, a comparison of allowed bounds on ($A_l,\ V_l$) depicted in Fig.~2
with those obtained from a similar analysis of 
$\sigma_{\mu\mu}$ and $A_{\rm FB}$ indicates that the new observables 
$\sigma_+$ and $\sigma_-$ have a slightly better sensitivity. This reflects 
the fact that, according to (\ref{deltasig+}), (\ref{Dsigm}) and
(\ref{coupl}), the deviations $\Delta\sigma_+$ and $\Delta\sigma_-$ are 
proportional to combinations of $Z^\prime$ coupling constants 
($V_l^2\pm A_l^2$, respectively) rather than the individual
ones, so that a one-parameter fit applies in this case, rather than a 
two-parameter one. Moreover, we may remark that the above analysis is a 
model-independent one, as it starts from rather general considerations. 
Therefore, it can usefully complement the conventional analysis of 
$Z^\prime$ couplings based on the observables $\sigma_{\mu\mu}$ and 
$A_{\rm FB}$.

\section{Effects of the anomalous gauge couplings}
In this regard, we now discuss a specific case of models with anomalous 
gauge couplings. This case was considered in Ref.~\cite{Hagiwara}, 
specifically concentrating on the $SU(2)\times U(1)$ invariant 
Lagrangian with only dimension six operators. 
Of these operators, four ones (denoted as ${\cal O}_{BW}$, ${\cal O}_{DW}$,
${\cal O}_{DB}$, and ${\cal O}_{\Phi,1}$) affect the neutral current 
and charged current amplitudes at tree level, and five more  
induce effects at the one-loop level. Of these five operators, three ones  
lead to anomalous trilinear vector boson couplings. 
As it was shown in Ref.~\cite{Hagiwara}, the quadratically and logarithmically 
divergent contributions of the latter five operators to neutral and 
charged current amplitudes are equivalent to a renormalization of the four
operators ${\cal O}_{BW}$, ${\cal O}_{DW}$, ${\cal O}_{DB}$, 
and ${\cal O}_{\Phi,1}$ which contribute at tree level. 
Although the number of parameters of this model is apparently rather large, 
it was shown in Ref.~\cite{Verzegnassi} that in the $Z$-peak subtracted 
representation only two parameters remain in the process $e^+e^-\to \bar ff$, 
since $f^r_{BW}$ and $f^r_{\Phi,1}$ are fully reabsorbed in the input 
parameters $\Gamma^{ll}_Z$ and $s_W^2(M_Z^2)$.
\par
In the $Z$-peak subtracted representation, the observables
$\sigma_{\mu\mu}$ and $\sigma_{\rm FB}$ are given by \cite{Verzegnassi}:

\begin{eqnarray}
\sigma_{\mu \mu}(s) &=&  \sigma_{\mu \mu}^{(\gamma )} (s) + \sigma_{\mu
\mu}^{(Z)}(s) + \sigma_{\mu \mu}^{(\gamma
Z)}(s), \label{sigma}\\
\sigma_{\mu \mu}^{(\gamma )} (s) & = &\tilde\sigma_{pt}
\left [ 1 + 2\tilde{\Delta} \alpha (s) \right ],
\label{sigma_gamma}\\
\sigma_{\mu \mu}^{(Z)} (s)&=&\tilde\sigma_{pt} \hskip 2pt \left | \chi_Z
\right |^2 a^4\hskip 2pt
\left [ 1 -2 R(s) - 16 s_W c_W \frac{\tilde{v}_l}{1+\tilde{v}^2_l}
V(s) \right ] , \label{sigma_Z}\\
\sigma_{\mu \mu}^{(\gamma Z)} (s)&=&\tilde\sigma_{pt} \hskip 2pt 2 
{\rm Re} \chi_Z \hskip 2pt
a^2\frac{\tilde{v}^2_l}{1+\tilde{v}^2_l} \hskip 3pt \times \nonumber \\
&\times& \hskip 3pt \left [ 1 + \tilde{\Delta }\alpha (s) - R(s) - 8
\hskip 2pt
\frac{s_W c_W}{\tilde{v_l}} V(s)
\right ], \label{sigma_gammaZ}
\end{eqnarray}
and
\begin{eqnarray}
\sigma_{\rm FB}(s)&=&\sigma_{\rm FB}^{(Z)}(s) + 
\sigma_{\rm FB}^{(\gamma Z)}(s), \label{sigma_FB}\\
\sigma_{\rm FB}^{(Z)}(s)&=&\frac{3}{4}\tilde\sigma_{pt}\left | \chi_Z 
\right |^2 
a^4\frac{4\tilde{v}_l^2}{(1+\tilde{v}^2_l)^2}
\left [ 1 - 2 R(s) - 8  \frac{s_W c_W}{\tilde{v_l}}
V(s) \right ] , \label{FB_Z}\\
\sigma_{\rm FB}^{(\gamma Z)}(s) & = &\frac{3}{4}\tilde\sigma_{pt} \hskip 2pt 
2{\rm Re} \chi_Z \hskip 2pt a^2\frac{1}{1+\tilde{v}^2_l}
\hskip 2pt \left [ 1 + \tilde{\Delta }\alpha (s) - R(s) \right ] .
\label{FB_gammaZ}
\end{eqnarray}
Here: ${\displaystyle{\tilde\sigma_{pt}=\frac{4 \pi \alpha^2}{3 s}}}$,
${\displaystyle{a^2=\frac{3 \Gamma^{ll}_Z}{\alpha M_Z}}}$,
$\alpha \equiv \alpha(0) \simeq 1/137$,
$s^2_W\equiv\sin^2\theta_W^{eff}$ as in (\ref{impborn}), 
and $\tilde{v}=1-4s_W^2$.
\par
Contributions of anomalous gauge couplings (AGC) to the above equations 
can be written as:
\begin{eqnarray}
&&\tilde{\Delta}^{(AGC)} \alpha (s) = - 8 \pi \alpha \frac{s}{\Lambda^2}
\left [
f^r_{DW} + f^r_{DB} \right ] ,
\label{alpha} \\
&&R^{(AGC)} (s) = \frac{8\pi \alpha }{s^2_W c^2_W} \hskip 2pt \frac{(s -
M^2_Z)}{\Lambda^2} \left [ c^4_W f^r_{DW} +
s^4_W f^r_{DB} \right ] ,  \label{R}\\
&&V^{(AGC)} (s) = \frac{8\pi \alpha }{s_W c_W} \hskip 2pt \frac{(s -
M^2_Z)}{\Lambda^2} \left [ c^2_W f^r_{DW} - s^2_W
f^r_{DB} \right ]. \label{V}
\end{eqnarray}
In these equations, $\Lambda$ is the mass scale of the new interaction, and
$f^r_{DW}$ and $f^r_{DB}$ are the renormalized couplings that are 
associated to the six-dimensional effective operators 
\begin{eqnarray}
&&{\cal O}_{DW}  = Tr \hskip 2pt \left(\left[ D_\mu , \hat{W}_{\nu \rho} 
\right]\left[ D^\mu , \hat{W}^{\nu \rho} \right ] \right ), \\
&&{\cal O}_{DB}  =  -\frac{g'^2}{2} \left (\partial_\mu B_{\nu \rho} \right)
\left(\partial^\nu B^{\nu \rho}\right),
\end{eqnarray}
that are involved in the effective Lagrangian \cite{Hagiwara} and survive in 
the $Z$-peak subtracted representation.
\par 
It is convenient to express the deviations of the cross section and of the 
forward-backward cross section from the SM prediction, at energies away from 
the $Z$ peak, in the approximation based on the simplified assumption 
{\it ii}), and neglecting numerically irrelevant small contributions. 
The result is:  
\begin{eqnarray}
\Delta\sigma_{\mu\mu}\approx \sigma_{pt}
\left(-8\pi\alpha\frac{s}{\Lambda^2}\right)
\left[\left(2.4\ f^r_{DW}+1.8\ f^r_{DB}\right)
+a^2{\rm Re}\chi_Z\left(2.4\ f^r_{DW}+0.2\ f^r_{DB}\right)\right]
\label{crossagc}
\end{eqnarray}
and
\begin{eqnarray}
\Delta\sigma_{{\rm FB}}\approx\frac{3}{4}\sigma_{pt}
\left(-8\pi\alpha\frac{s}{\Lambda^2}\right)
\left[\left(2.4\ f^r_{DW}+0.2\ f^r_{DB}\right)
+a^2{\rm Re}\chi_Z\left(2.4\ f^r_{DW}+1.8\ f^r_{DB}\right)\right].
\label{fbagc}
\end{eqnarray}
The corresponding expressions for the new 
observables $\sigma_+$ and $\sigma_-$ can be easily derived  
from the relation (\ref{relation}) and the formulae for 
$\sigma_{\mu\mu}$ and $\sigma_{\rm FB}$. The result is:
\begin{eqnarray}
\Delta\sigma_+\approx \sigma_{pt}
\left(-8\pi\alpha\frac{s}{\Lambda^2}\right)
\left(1+a^2{\rm Re}\chi_Z\right)
\left(2.4\ f^r_{DW}+1.0\ f^r_{DB}\right),
\label{sigma+agc}
\end{eqnarray}
and
\begin{eqnarray}
\Delta\sigma_-\approx \sigma_{pt}
\left(-8\pi\alpha\frac{s}{\Lambda^2}\right)
\left(1-a^2{\rm Re}\chi_Z\right)\ 0.8\ f^r_{DB}.
\label{sigma-agc}
\end{eqnarray}
\par 
Fig.~3a shows the relative deviation $\Delta\sigma_{+}/\sigma_{+}^{SM}$ 
induced by the anomalous gauge couplings with ($f^r_{DW},\ f^r_{DB}$)=
$(\pm 1,\pm 1)$ and $\Lambda=1$ TeV, as numerical examples. In contrast to the 
$Z^\prime$ case (Fig.~1a), the sign of the deviation due to anomalous 
gauge couplings is not defined, i.e., it may be either positive or negative 
at energies up to LEP2. However, as Eq.~(\ref{sigma+agc}) shows, characteristic
energy correlations of the deviation occur also in this case. Indeed, for 
certain values of the parameters ($f^r_{DW},\ f^r_{DB}=(1,1); (1;-1)$)  
the energy behaviour of the deviation due to anomalous gauge couplings is 
the same as that in Fig.~1a for the $Z^\prime$, while for other choices of the 
parameters ($f^r_{DW},\ f^r_{DB}=(-1,1); (-1;-1)$) the behaviour is opposite 
to the $Z^\prime$ case. 
\par 
Fig.~3b shows the energy behaviour of $\Delta\sigma_-/\sigma_-^{SM}$ for the 
same choice of the parameters. Such behaviour is qualitatively similar to the 
$Z^\prime$ case, in particular regarding the energy correlations of the 
sign of the deviation. Moreover, from Eq.~(\ref{sigma-agc}), $\Delta\sigma_-$ 
depends on the couplings $f^r_{DB}$ only, and therefore just a 
one-parameter fit to current (and expected) experimental data is required 
for this parameter in our analysis.  
\par 
Concerning the distinction of the two potential sources of nonstandard 
effects considered here, $Z^\prime$ vs anomalous gauge couplings, one can 
use the negative sign property at LEP2 of $\Delta\sigma_+$ as well as
$\Delta\sigma_{\mu\mu}$ and $\Delta\sigma_{\rm FB}$, which holds for the 
former case (Table 1) but not for the latter one. The hatched area in the 
($f^r_{DW}$,$f^r_{DB}$) plane of Fig.~4 corresponds to the values of 
anomalous gauge couplings such that the deviations of
Eqs.~(\ref{crossagc})-(\ref{sigma+agc}) are negative at LEP2. This is 
a `confusion' region, where both nonstandard mechanisms can induce the same 
deviations from the SM 
prediction, and therefore cannot be distinguished through the property of 
negative sign deviation of the observables
introduced here. On the contrary, the white area on the left of the hatched 
one is determined by values of the anomalous couplings such that the 
corresponding (positive) deviations are not allowed to the $Z^\prime$ and, in 
this sense, the source of the nonstandard effect can be distinguished. 
\par 
Another condition can be derived by considering that Eqs.~(\ref{deltasig+}) and 
(\ref{Dsigm}) imply, for the $Z^\prime$, that 
\begin{equation}
\frac{\Delta\sigma_-}{\Delta\sigma_+}=\frac{1-a_l^2{\rm Re}\chi_Z}
{1+a_l^2{\rm Re}\chi_Z}\cdot\frac{v_l^{\prime 2}-a_l^{\prime 2}}{v_l^{\prime
2}+a_l^{\prime 2}}, \label{zprime}\end{equation}
with, of course, $\vert (v_l^{\prime 2}-a_l^{\prime 2})/(v_l^{\prime
2}+a_l^{\prime 2})\vert\le 1$. For, e.g., $\sqrt s=190$ GeV, the above ratio 
is limited by
\begin{equation}
\vert\frac{\Delta\sigma_-}{\Delta\sigma_+}\vert\le 0.36.\label{ratio}
\end{equation}
Taking into account Eqs.~(\ref{sigma+agc}) and (\ref{sigma-agc}) 
for the case of the anomalous gauge couplings: 
\begin{equation} 
\frac{\Delta\sigma_-}{\Delta\sigma_+}=\frac{1-a_l^2{\rm Re}\chi_Z}
{1+a_l^2{\rm Re}\chi_Z}\cdot\frac{0.8 f^r_{DB}}{2.4f^r_{DW}+1.0f^r_{DB}}, 
\label{ragc}\end{equation}
and, comparing with the numerical bound (\ref{ratio}), one finds that 
the actual `confusion' region between the two kinds of mechanism is limited by 
the lines 4 and 5 of Fig.~4, and therefore is somewhat reduced.
\par
Finally, the contours in Fig.~5 represent the upper bounds on $f^r_{DW}$ and 
$f^r_{DB}$ allowed by the combination of $\sigma_+$ and $\sigma_-$ at the 
2-$\sigma$ level, in the case of no deviation observed from the SM, using the 
current LEP2 data of Table 2 and the expected experimental uncertainty with 
the future full luminosity. Numerical values of the anomalous gauge
couplings included in the area between the two ellipses give rise to 
deviations from the SM that are, potentially, still in the reach of future 
measurements at LEP2. The values of the anomalous couplings within the 
shaded area produce deviations which can also originate from the $Z^\prime$, 
and in this sense identify the so-called `confusion' area.   

\section*{Acknowledgements}
AAB and AAP gratefully acknowledge the support of the University of Trieste.

\section*{Figure captions}
\begin{description}

\item{\bf Fig.~1a} 
Relative deviation $\Delta\sigma_{+}/\sigma_{+}^{SM}$ induced by a 
`sequential' $Z^\prime$ with $v^\prime_l=v_l$ and $a^\prime_l=a_l$.
$M_{Z^\prime}=500~{\rm GeV}$ (line 1) and $M_{Z^\prime}=700~{\rm GeV}$ 
(line 2). Notice $\Delta\sigma_{+}/\sigma_{+}^{SM}=0$ at
$\sqrt{s}={M_Z}/{\sqrt{1+a_l^2}}\simeq78~{\rm GeV}$
and $\sqrt{s}={M_Z}$.
\item{\bf Fig.~1b} 
Relative deviation $\Delta\sigma_{-}/\sigma_{-}^{SM}$.
Lines 1 and 2 correspond to ${{v^\prime}_l}^2-{{a^\prime}_l}^2=
v_l^2-a_l^2$, $M_{Z^\prime}=500~{\rm GeV}$ and 
$M_{Z^\prime}=700~{\rm GeV}$, respectively.
Lines 3 and 4 correspond to ${{v^\prime}_l}^2-{{a^\prime}_l}^2=
-(v_l^2-a_l^2)$ and $M_{Z^\prime}=500~{\rm GeV}$, 
$M_{Z^\prime}=700~{\rm GeV}$, 
respectively. Notice $\Delta\sigma_{-}/\sigma_{-}^{SM}=0$ at $\sqrt{s}={M_Z}$ 
and $\sqrt{s}={M_Z}/{\sqrt{1-a_l^2}}\simeq113~{\rm{GeV}}$.
\item{\bf Fig.~2} 
Upper bounds on the $Z^\prime$ couplings ($A_l$, $V_l$)
determined by $\sigma_\pm$ at the $2\sigma$ level. The solid contour is 
obtained from $e^+e^-\to\mu^+\mu^-$ with inputs summarized in Table~2. The  
dashed line is obtained with $L_{int}=500~\mbox{pb}^{-1}$ 
and $\sqrt{s}=190\ {\rm GeV}$.
\item{\bf Fig.~3a} 
Relative deviation $\Delta\sigma_{+}/\sigma_{+}^{SM}$ induced by anomalous 
gauge couplings. Lines $1,2,3$ and 4 correspond to:  
($f^r_{DW},\ f^r_{DB}$)=$(1,1)$; $(1,-1)$; 
$(-1,1)$; $(-1,-1)$, respectively.
\item{\bf Fig.~3b} Relative deviation of $\Delta\sigma_{-}/\sigma_{-}^{SM}$,  
parameters as in Fig.~3a.
\item{\bf Fig.~4} 
Straight lines in the two-dimensional plane 
($f^r_{DW},\ f^r_{DB}$) corresponding to $\Delta\sigma_{\mu\mu}=0$ (line 1), 
$\Delta\sigma_{+}=0$ (line 2) and $\Delta\sigma_{\rm FB}=0$ (line 3).
The hatched area is the `confusion' area between $Z^\prime$ and anomalous
gauge couplings effects when the deviations $\Delta\sigma_{\mu\mu}$, 
$\Delta\sigma_{\rm FB}$ and $\Delta\sigma_+$ are negative at LEP2. 
The shaded area is the reduced `confusion' area using also the solutions of  
(\ref{ratio}) and (\ref{ragc}) (lines 4 and 5).
\item{\bf Fig.~5} 
Upper bounds on the anomalous gauge couplings ($f^r_{DW},\ f^r_{DB}$)
determined by $\sigma_\pm$ at the $2\sigma$ level. 
Input parameters for solid and dotted lines are as in Fig.~2.
The shaded area is the `confusion' one, where $Z^\prime$ and anomalous gauge 
couplings can produce the same deviation.
\end{description}

\vfill\eject
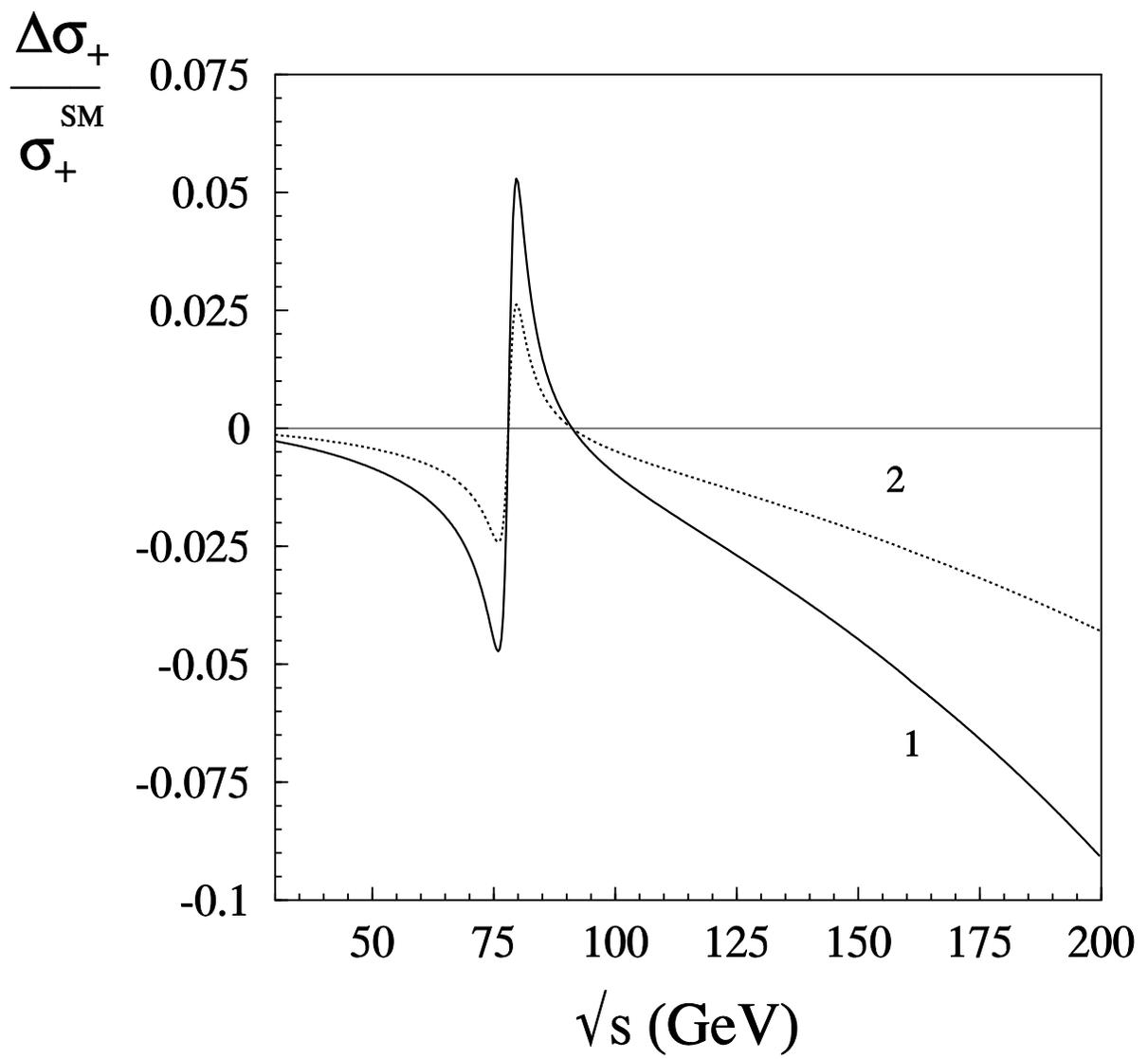
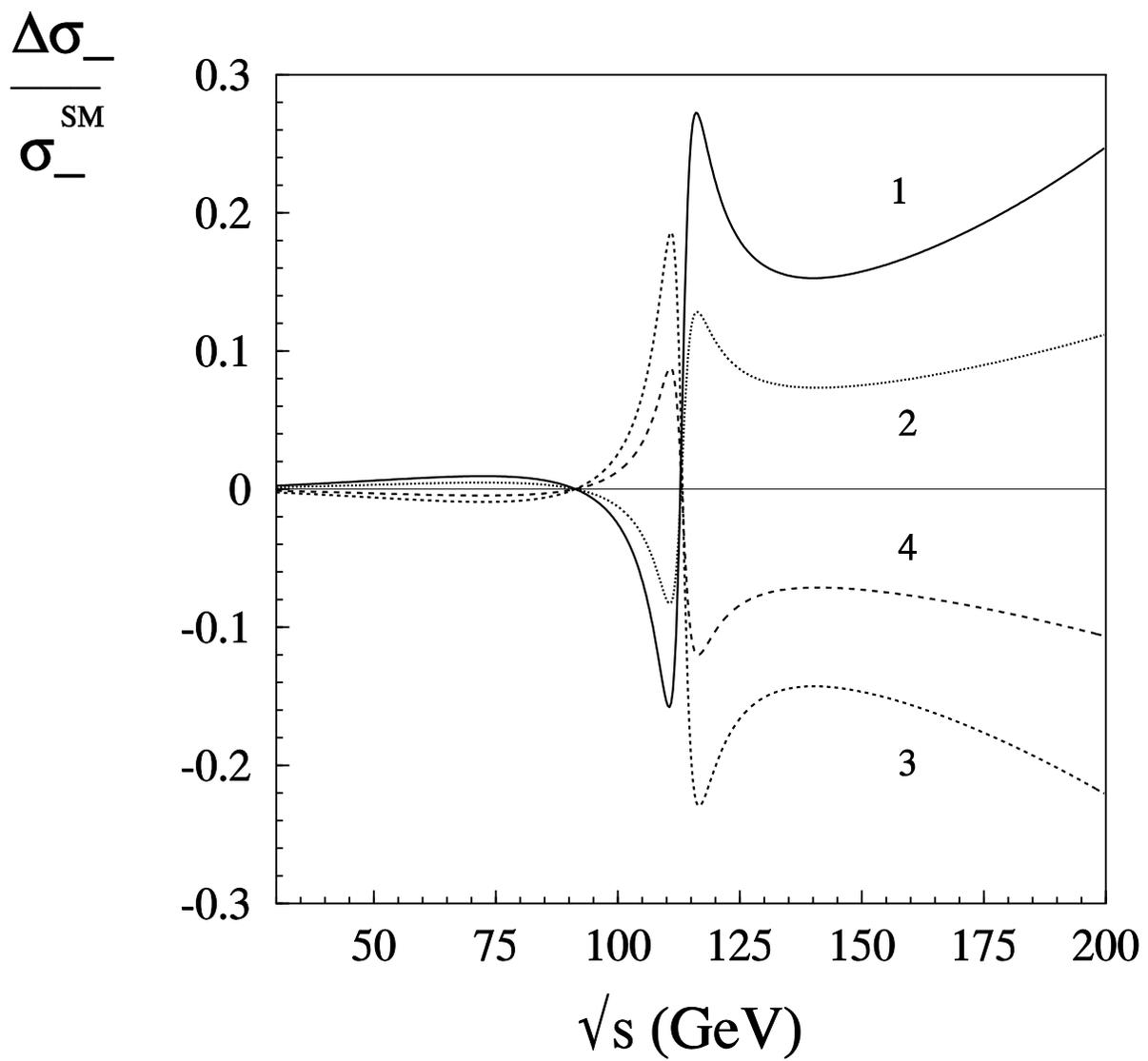
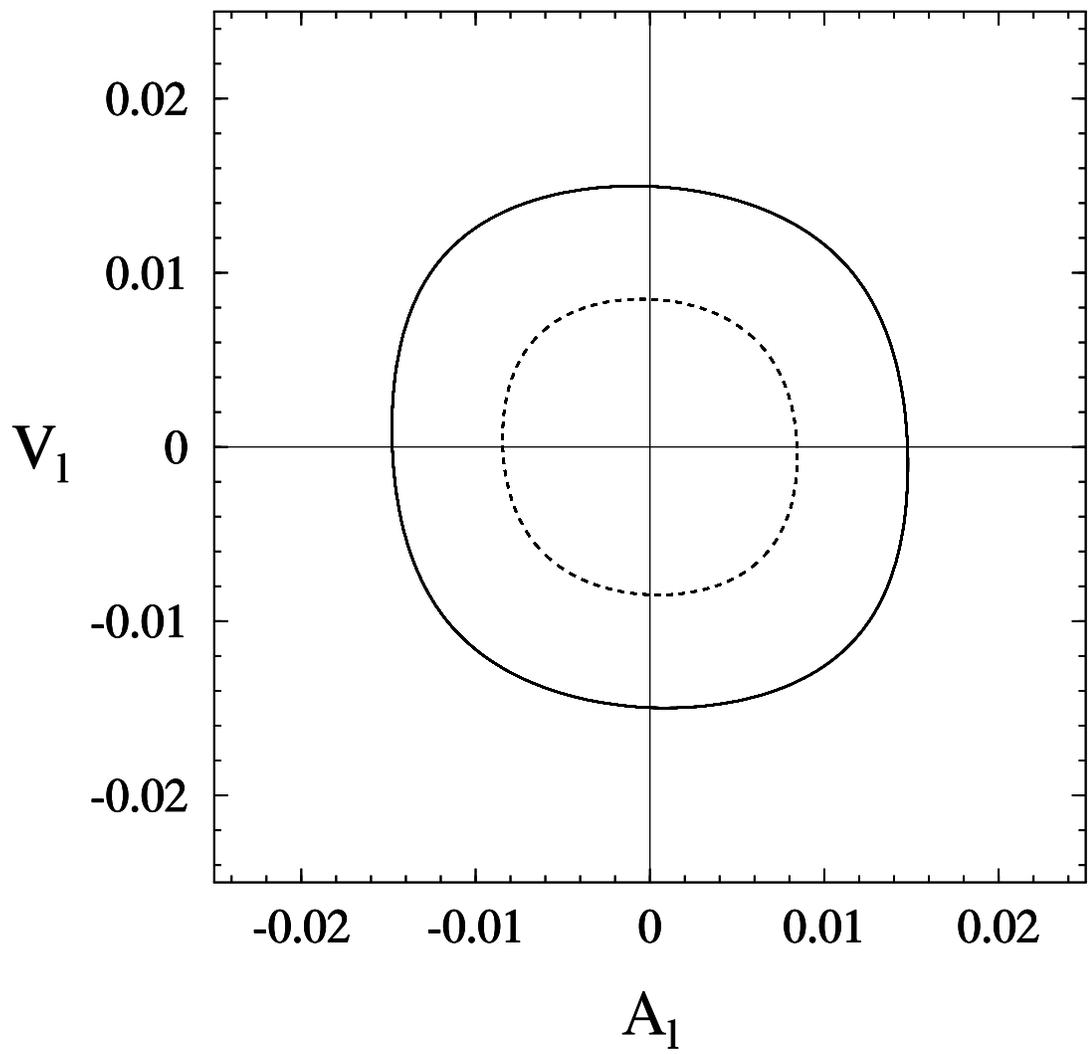
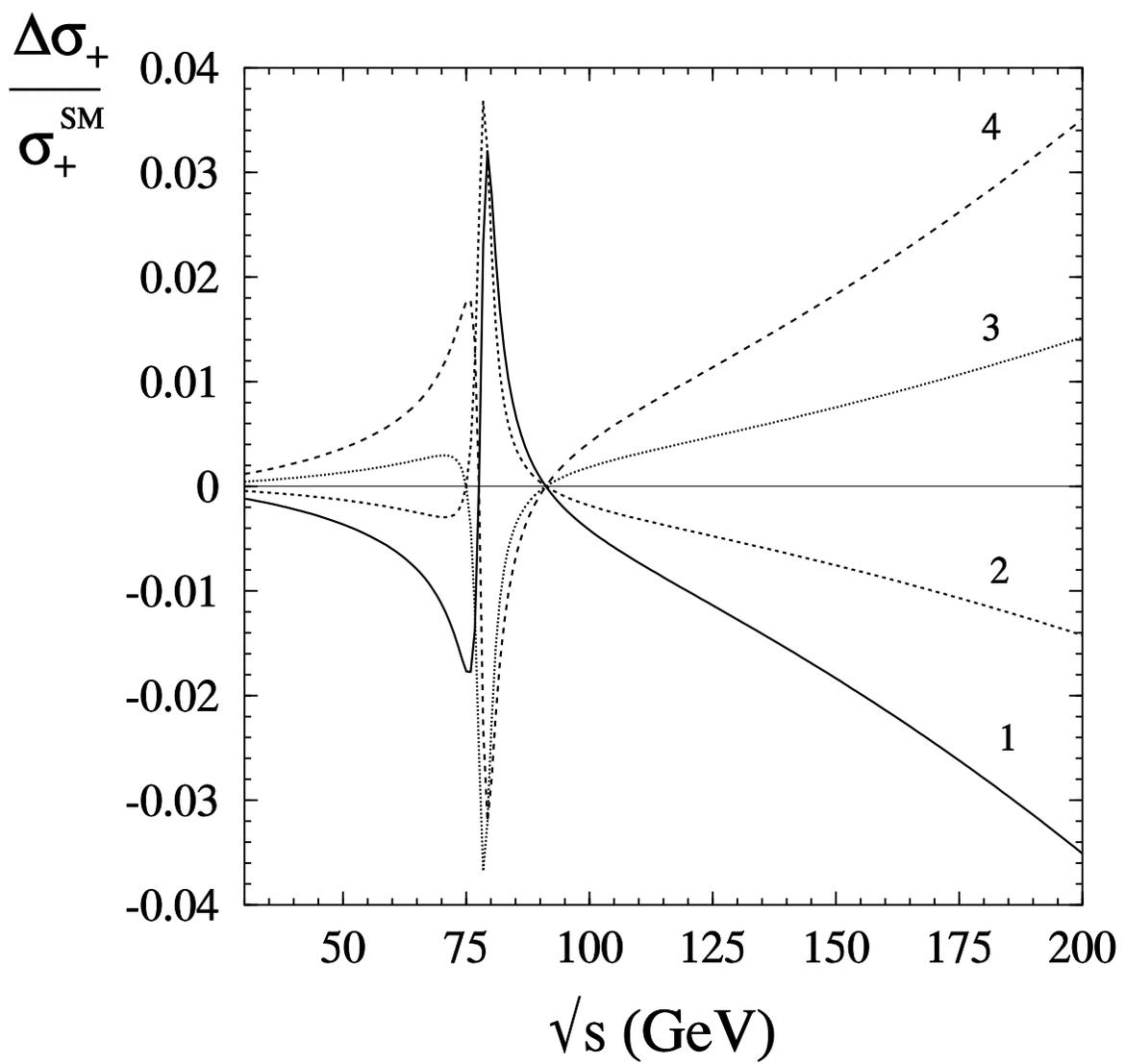
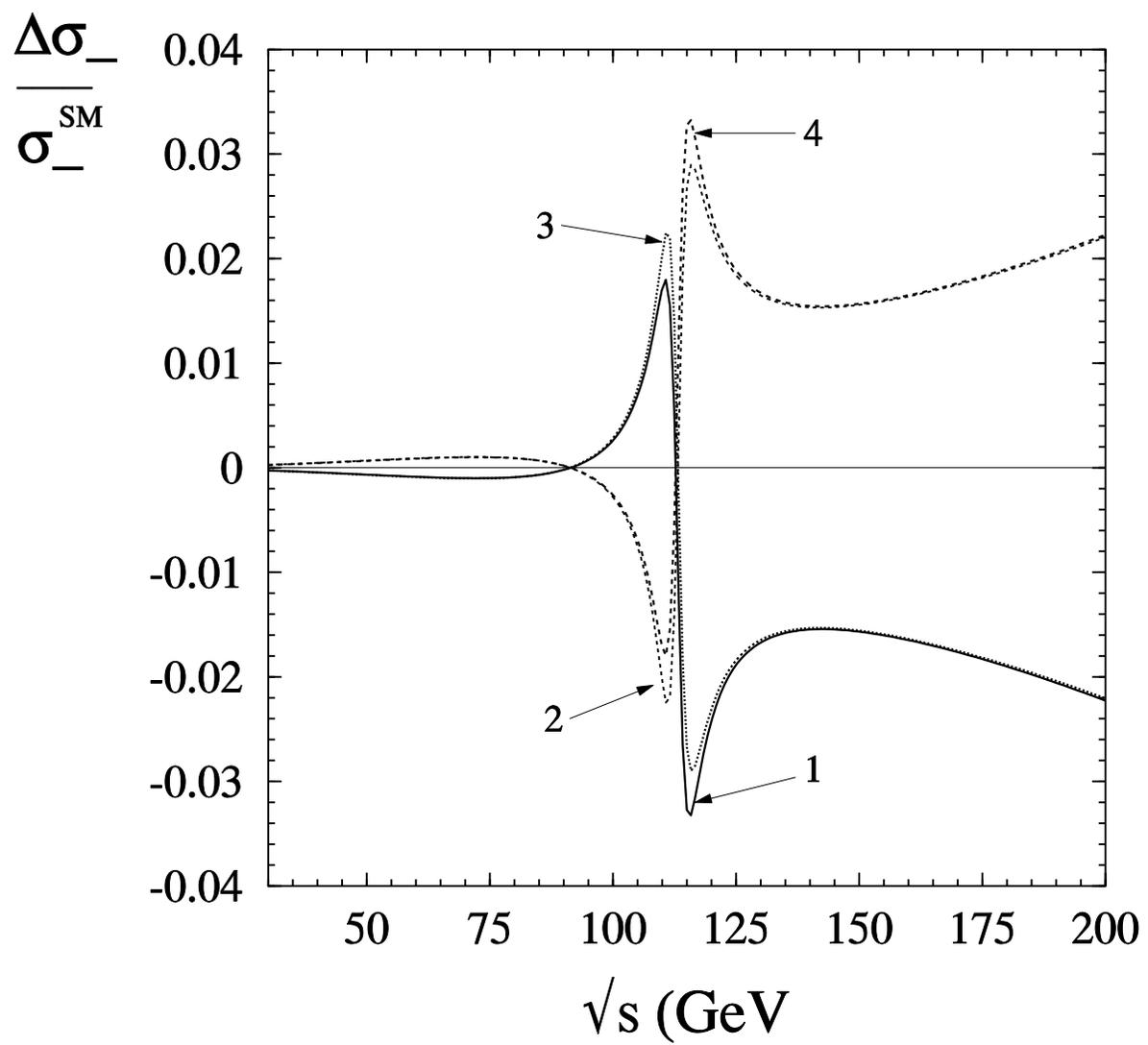
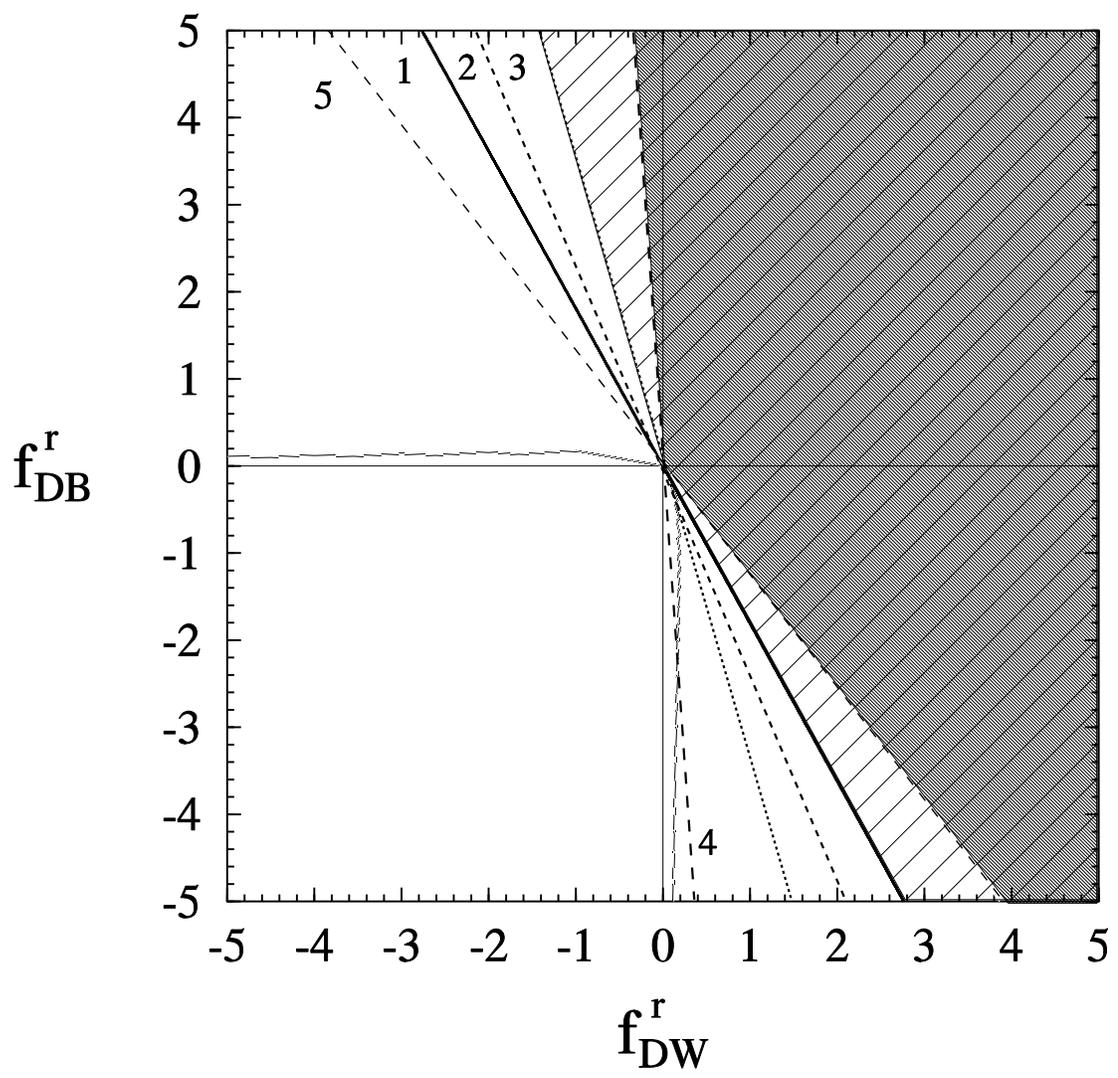
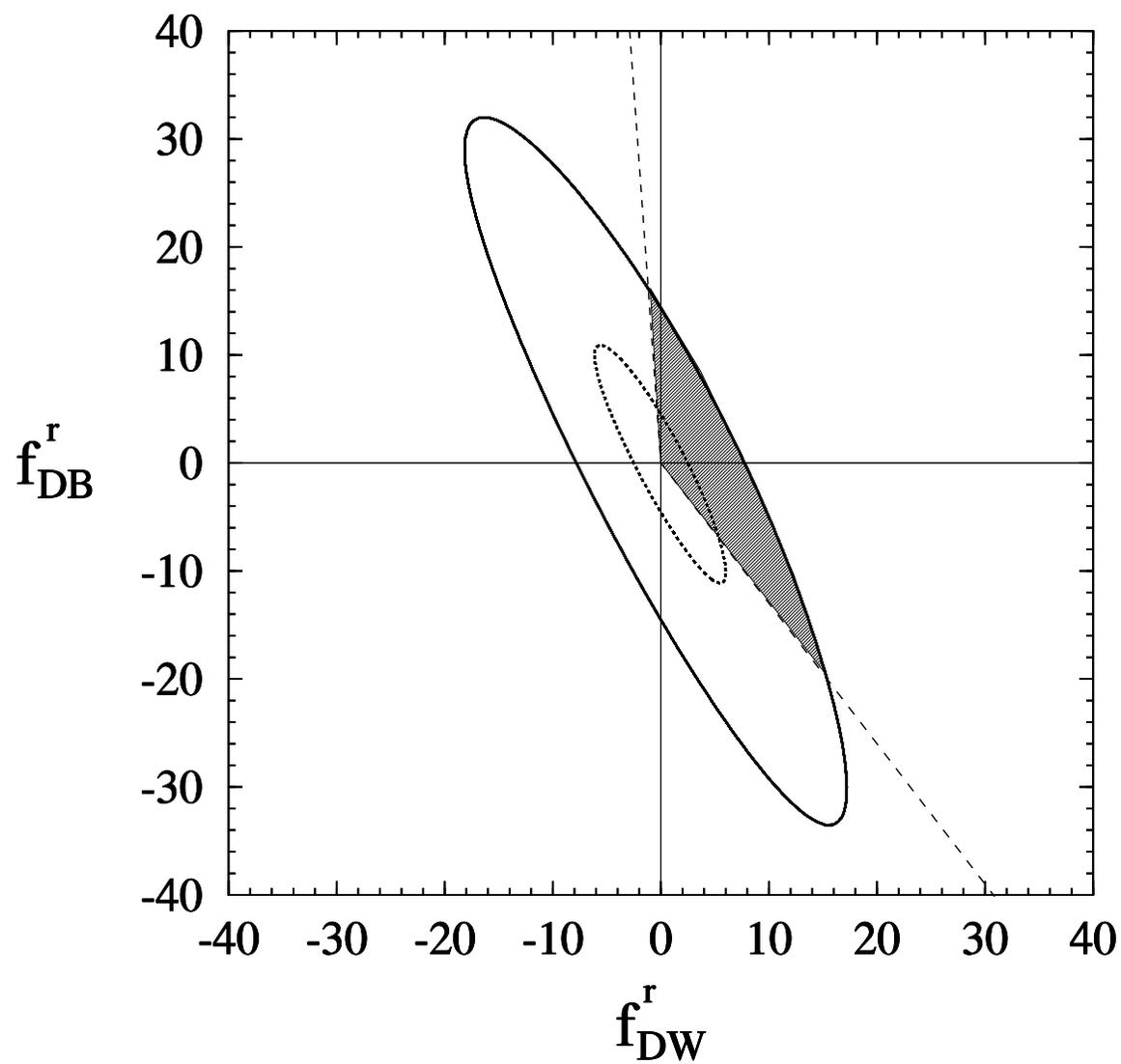
\end{document}